\documentclass[12pt,preprint]{aastex}

\slugcomment{}
\shorttitle{}
\shortauthors{}

\begin{document}
\title{Fabry-P\'erot observations of the HH~110 jet}
\author{A. Riera
\altaffilmark{1,2,3} 
}
\affil{Departament de F\'\i sica i Enginyeria Nuclear, Universitat Polit\'ecnica de Catalunya, Av. V\'\i ctor 
Balaguers/n E-08800 Vilanova i La Geltr\'u, Spain }
\affil{Departament d'Astronomia i Meteorologia, Universitat de Barcelona, Av. Diagonal 647, E-08028 Barcelona, Spain}
\affil{On sabbatical leave at the Instituto de Ciencias Nucleares, UNAM}
\email{angels.riera@upc.es,ariera@nuclecu.unam.mx}

\author{A. C. Raga
\altaffilmark{3}}
\affil{Instituto de Ciencias Nucleares, UNAM, Ap. 70-543, 04510 D.F., M\'exico}

\email{raga@astroscu.unam.mx}

\author{B. Reipurth
\altaffilmark{4}}
\affil{Institute for Astronomy, University of Hawaii, 2680 Woodlawn Drive, Honolulu, HI 96822, USA }

\email{reipurth@ifa.hawaii.edu}

\author{Ph. Amram, J. Boulesteix  
\altaffilmark{5}}
\affil{Laboratoire d'Astrophysique de Marseille, 2 Place Le Verrier, 13248 Marseille, France}

\email{philippe.amram@oamp.fr,jacques.boulesteix@oamp.fr}

\author{J. Cant\'o, O. Toledano 
\altaffilmark{6}}
\affil{Instituto de Astronomia, UNAM, Ap. 70-264, 04510 D.F., M\'exico}

\begin{abstract}
We have obtained a H$\alpha$ position-velocity cube from Fabry-P\'erot
interferometric observations of the HH~110 flow. We analyze the results
in terms of anisotropic wavelet transforms, from which we derive
the spatial distribution of the knots as well as their characteristic
sizes (along and across the outflow axis). We then study the spatial
behaviour of the line width and the central radial velocity. The
results are interpreted in terms of
a simple ``mean flow+turbulent eddy'' jet/wake model. 
We find that most of the observed kinematics appear to be a
direct result of the mean flow, on which are superposed low
amplitude ($\sim 35$~km~s$^{-1}$) turbulent velocities.
\end{abstract}



\section{Introduction}

The HH 110 Herbig-Haro jet was discovered by Reipurth \& Olberg (1991), and it presents a complex structure of aligned knots, 
which opens out with an unusually large angle of $\sim 10^\circ$.  While HH objects such as HH 34 or HH 111 show aligned 
knots with more or less organized, arc-like shapes, HH 110 has a more chaotic structure, more reminiscent of a turbulent flow. 

A possible explanation for the fact that HH 110 is different from other HH jets has been proposed by 
Reipurth, Raga \& Heathcote (1996), who suggested that HH 110 is the result of a deflection of the 
adjacent HH 270 jet (which travels approximately in an E-W direction) through a collision with a dense molecular cloud core.
This interpretation was supported by the fact that while the source of the HH 270 jet is detected in the radio 
free-free continuum, no such source is detected along the HH 110 axis (Rodr\'\i guez et al.  1998).  Also, infrared 
images show H$_2$ emission concentrated on the W side of the HH 110 jet (Davis, Mundt \& Eisl\"offel 1994; Noriega-Crespo et al. 1996), 
which would correspond to the region in which the jet is in contact with the dense cloud which produces the deflection.  
Against the jet/cloud collision scenario are the results of Choi (2001), who do not find molecular (HCO$^+$) emission around 
the jet/cloud impact point, as might be expected. 

There are a series of papers which describe analytic and numerical models meant specifically for HH 110 as a jet/cloud impact 
flow (Raga \& Cant\'o 1995; de Gouveia dal Pino 1999; Hurka, Schmid-Burgk \& Hardee 1999).  The latest models of this object are the 
ones of Raga et al. (2002), who have carried out the most detailed
simulations up to now of this object, and computed H$\alpha$ and molecular hydrogen emission maps which agree with the 
observations in a qualitatively successful way.  These authors also suggest that the point at which the HH 270 jet is 
currently impacting the cloud might have moved westwards of the HH 110 axis (i. e., deeper into the cloud), 
which would explain the lack of HCO$^+$ emission in the crossing region of the HH 270 and HH 110 axes. 

In a recent paper, Riera et al. (2003) have discussed the kinematics and excitation along and across the HH 110 outflow, 
as derived from long-slit spectra obtained with 5 slit positions (one along and four across the outflow axis), finding 
very complex structures.  In the present paper, we discuss new Fabry-P\'erot data which cover only the H$\alpha$ line, 
but give a full spatial coverage of the well resolved HH 110 outflow.  The full spatial coverage of the position-velocity 
cube obtained from our data allows us to carry out a somewhat novel analysis (at least, in the context of HH objects).  
As the observed spatial and kinematical structures are very chaotic, and appear to defy a direct description, we have 
chosen to describe these structures through a wavelet analysis.  A wavelet transform of an image gives information 
similar to the one that could be obtained from a more standard Fourier analysis, but also preserves the spatial information. 

In this way one can obtain, for example, the characteristic sizes (along and across the outflow axis) of the knots as a 
function of position along the jet.  This kind of information could also be retrieved by carrying out fits to the intensity 
profiles of the knots (e. g., with anisotropic Gaussian functions, as has been done in the past for other objects by 
Raga, Mundt \& Ray 1991), but such an approach is impractical for an object as complex as HH 110, in which knots of different 
shapes coexist across the sections of the outflow. 

We therefore obtain a position-velocity cube from our Fabry-P\'erot observations (described in \S2), and obtain velocity 
channel maps from which we compute an H$\alpha$ image, as well as radial velocity and line width maps (see \S3).  
We then carry out an anisotropic wavelet analysis of the H$\alpha$ image, and determine the spatial scales present in the
 map as a function of position along the outflow (\S4).  We then study the variations of the line width and the radial 
velocity of the emission as a function of spatial scale and of position along and across the jet (\S5).  Finally, we 
interpret this kinematic information in terms of a ``mean flow+turbulent eddy'' model of a turbulent jet (or wake), 
and use the model to deduce the mean flow parameters that are relevant for the HH 110 jet (\S6).  The results are 
then summarized in $\S7$.

\section{ Observations}

The observations were obtained at the
European Southern Observatory 3.6m telescope  in January 1997.  The Fabry-P\'erot instrument CIGALE \footnote{The instrument 
CIGALE (for CIn\'ematique des GALaxieEs) is a visiting instrument belonging to the Laboratoire d'Astrophysique de Marseille.} 
was used.  It is composed of a focal reducer (bringing the original f/8 focal ratio of the Cassegrain focus to f/2), a 
scanning Fabry-P\'erot and an Image Photon Counting System (IPCS).  Table 1 contains the journal of the observations.
We have used a Fabry-P\'erot with an interference order $p = 793$ (380 km s$^{-1}$) and a {\it Finesse} $F =11$.
The IPCS, with a time sampling of 1/50 second and zero readout noise, makes it possible to scan the interferometer rapidly, 
avoiding sky transparency, air-mass and seeing variation problems during the exposures and thus has several advantages over 
a CCD for this application.  The exposure time per elementary channel was 5 seconds.  Several observations of HH~110 have been 
obtained in H$\alpha$ during 2 nights.
The two sets of observations in the H$\alpha$ line have been done using
2 different interference filters in order to check if the shape of the
interference filter affects the shape of the H$\alpha$ profiles.
A $18300$~s exposure was obtained with a $\lambda_0=6563$~\AA\
(central wavelength) and $\Delta \lambda=9$~\AA\ (FWHM of transmission
function) filter, and a $3600$~s exposure was obtained
with a $\lambda_0=6564$~\AA\ and $\Delta \lambda=17$~\AA\ filter.

The reduction of the data cubes was performed using the CIGALE software. 
The data reduction procedure has been extensively described in
Amram et al. (1992, 1993, 1994, 1995, 1996, 1998).  Wavelength
calibration was obtained by scanning the narrow Ne 6599 \AA\ line under the same conditions as the observations.  
Velocities measured relative to the systemic velocity are very accurate, with an error of a fraction of a channel width 
($<$ 3 km s$^{-1}$) over the whole field. 

The signal measured along the scanning sequence was separated into two parts: (1) an almost constant level produced by the 
continuum light in a narrow passband around the line and (2) a varying part produced by the line.
The continuum level was taken to be the mean of the three faintest channels, to avoid channel noise effects. The integrated 
line flux map was obtained
by integrating the monochromatic profile in each pixel.  The velocity sampling was 16 km s$^{-1}$. 
Strong OH night sky lines passing through the filters were subtracted by determining the level of emission from 
extended regions away from the HH object (Laval et al. 1987). 

\section{ The H$\alpha$ channel maps, radial velocities and line widths}

Figure 1 shows the velocity channel maps obtained from the Fabry-P\'erot data of HH 110.  We detect H$\alpha$ emission in 
heliocentric radial velocities ranging from $-100$ to +80 km s$^{-1}$, with peak intensities in the $-38.3$ and $-22.5$ km s$^{-1}$ 
channel maps. 

From the line profiles at each position, we have computed the velocity of the emission peak (with a parabolic fit) and the 
FWHM (with linear fits to the sides of the line profiles), as well as the line flux (by adding the fluxes of the individual 
channels).  The results are shown in Figure 2. 

The integrated H$\alpha$ map obtained from the present data looks similar to previously published H$\alpha$ images of this object, 
showing a complex structure of knots that broadens out to the South.  The radial velocities have low values 
($\sim 0 \to -60$ km s$^{-1}$), with more negative values to the South (see Fig. 2).  The radial velocity structure 
across the jet is complex, showing variabilities which do not appear to show systematic trends.  These results are 
qualitatively consistent with the radial velocities of this flow which have been previously obtained from long-slit 
spectroscopy (Reipurth et al. 1996; Riera et al. 2003).  The FWHM of the H$\alpha$ line ranges from 50 to 
100 km s$^{-1}$, and also shows a complex spatial structure (see Fig. 2). 

Because of the complexity of the observed structures, we have carried out an analysis based on wavelet transforms, 
which allows us to obtain a general description of the flow without attempting a somewhat hopeless ``knot by knot'' 
description of the observations. 

\section{Wavelet analysis of the H$\alpha$ image}

\subsection{The anisotropic wavelet transforms}

Wavelet transforms have been used in different astrophysical contexts. 
For example, Gill \& Henriksen (1990) have used wavelet transforms to
describe the complex structures observed in CO radial velocities of
molecular clouds. Obtaining a description of the complex structure of the
HH 110 jet belongs to the same class of problem, i.e., identifying
the characteristics of complex spatial structures without loosing the
positional information. Therefore, we have
applied a wavelet transform analysis to our Fabry-P\'erot data.
 
A 2D wavelet transform analysis of the H$\alpha$ image of HH~110
can be used to obtain the positions of all of the ``knots'', as
well as their characteristic sizes. This kind of procedure
is described, e.~g., in the book of Holschneider (1995).

Once we have obtained the positions and the characteristic sizes of the 
H$\alpha$ structures of the HH 110 jet from the wavelet spectrum,
we calculate the spatial averages of the deviations of the line center velocity
and if the line widths (averaging over the characteristic sizes of the
structures) in order to study the turbulence in the HH 110 jet (see \S 5).
This procedure is completely equivalent to the one followed by
Gill \& Henriksen (1990).

An interesting difference between our case and the ``turbulent
molecular cloud'' study of Gill \& Henriksen (1990) is that
a jet flow has two clearly defined directions~: along and
across the flow axis. Because of this, it is reasonable to expect
that the characteristics of the structures along and across the flow
axis might differ. To allow for this effect, we first rotate
the H$\alpha$ image so that the flow axis is parallel to the ordinate.
On this rotated image, we then carry out a decomposition
in a basis of anisotropic wavelets, which have different sizes
along and across the outflow axis.

In particular, we have used a basis of ``Mexican hat'' wavelets of the form
\begin{equation}
g(r) = C (2 - r^2) e^{-r^2/2} , 
\end{equation}
where $r = [ ( x/a_x)^2 + (y/a_y)^2]^{1/2}$, and $a_x$ and $a_y$ are the scale lengths of the wavelet along the $x$- 
and $y$-axes, respectively.  After trying several different possibilities, we have chosen a $C = (a_x^2 + a_y^2)^{-1/2}$ 
normalization. It is possible to use other normalizations (e.~g., $C=
\sqrt{a_x a_y}$), which lead to slightly different results, as the power
is rearranged over different regions of the wavelet spectrum. However, the
positions of the different peaks of the wavelet spectrum (see below) are
not strongly affected by the choice of normalization.

The process that is carried out in order to compute the wavelet transform is to choose a range for $a_x$ and $a_y$ 
(which are taken to have integer values of pixels).  For all of the possible $(a_x, a_y)$ pairs we then compute the convolutions 
\begin{equation}
T_{a_x, a_y} (x,y) = \int \, \int \, I (x^\prime, y^\prime) g(r^\prime) dx^\prime dy^\prime , 
\end{equation}
where $r^\prime = \{ [ ( x^\prime - x) / a_x]^2 + [(y^\prime - y)/a_y]^2 \}^{1/2}$, and $I (x,y)$ is the observed map as a function of the pixel position $(x,y)$.  These convolutions are calculated with a standard FFT algorithm. 

With this process, we then produce the transform maps $T_{a_x, a_y} (x,y)$, corresponding to versions of the observed map $I(x,y)$ which have been smoothed with the different $g_{a_x, a_y}$ wavelets.  These maps can be used as follows. 

We first fix a position $y$ along the observed jet.  Keeping this value
of $y$ fixed, we find the finite set of values $(x_k, a_{x,k}, a_{y,k})$
corresponding to the position across the jet and the spatial
scales (across and along the jet) at which the wavelet transform has
local maxima in the $a_x$ and $a_y$ dimensions as well as in either
the $x$ or in the $y$ dimensions. In this way, for each $y$ we detect
the positions $x_k$ of the
knots (which correspond to intensity maxima either along or across
the jet axis), as well as their characteristic sizes ($a_{x,k}$ and $a_{y,k}$).
In practice, at many positions $y$ along the jet one
finds a single maximum, but in some positions up to three maxima are
found, corresponding to different structures observed across the HH 110 jet. 

\subsection{The H$\alpha$ ``image''}

The results obtained with the process described in \S4.1 are shown in Figure 3.  This figure shows the H$\alpha$ image (see \S3), which we have rotated by 13$^\circ$ so that the outflow is more or less parallel to the $y$-axis.  We have then convolved this map with a set of wavelets with $1 \leq a_x \leq 20$ pixels (with $0^{\prime\prime}. 91$ per pixel) and $1\leq a_y \leq 20$ pixels.  The values of the peaks $x_k$ of the wavelet transform obtained for each position $y$ along the jet are shown in the right hand panel of Figure 3.  Also shown (as error bars) are the values of $a_{x,k}$ and $a_{y,k}$ corresponding to each peak, which give an estimate of the characteristic sizes (across and along the jet, respectively) of the observed structures. 

In order to quantify the observed broadening of the jet, for each position $y$ we compute the weighted mean of the $x$-spatial scale (perpendicular to the outflow axis): 
\begin{equation}
<a_x> = {\sum_k a_{x,k} T_{a_{x,k}, a_{y, k}} (x_k, y) \over \sum_k T_{a_{x, k}, a_{y, k}} (x_k, y) } , 
\end{equation}

Figure 4 shows $<a_x>$ 
as a function of position $y$ along the HH 110 flow.  In this figure, $y = 0$ corresponds to the position of the
peak of knot A, and $y$ grows moving South along the HH 110 axis.  From knot A to the middle of knot 
$\hbox{C} \, (y = 0 \to 20^{\prime\prime})$, 
$< a_x>$ grows from $\sim 3^{\prime\prime}$ to $\sim 7^{\prime\prime}$. 
For $y = 20^{\prime\prime} \to 70^{\prime\prime}, \, < a_x>$ increases 
more or less monotonically from $\sim 7^{\prime\prime}$ to $\sim 15^{\prime\prime}$,  
For $y = 70^{\prime\prime} \to 90^{\prime\prime}, \, < a_x>$ shows a more or less constant value with a small dispersion 
(with $< a_x>$ values ranging from  $\sim 10^{\prime\prime}$ to $\sim 13^{\prime\prime}$).
For $y > 90^{\prime\prime}$, $< a_x>$ shows a wide dispersion, with values ranging from $\sim 1^{\prime\prime}$ 
up to $\sim 15^{\prime\prime}$.

If we draw a straight line in Figure 4, passing through the peak $< a_x>$ of knots A and L, we see that it forms an upper 
envelope to all of the values of $<a_x>$ in this region (which includes all  knots from A to L).  
  From this upper envelope, we deduce that  the emitting region (from $y = 0^{\prime\prime} \to 70^{\prime\prime}$) broadens with 
a half-opening angle of 10$^\circ$.

Finally, in Figure 5 we show the $a_{y,k}$ as a function of $a_{x,k}$ for different regions along the HH 110 jet.
 We plotted together 
the points with values of $y$ in five different intervals along the jet beam (knots A-H, I-N, O-Q, 
R$_1$-R$_2$ and S, corresponding to the $y$-intervals given in Table~2).
  We find that for knots A-H, the $a_{y,k}$ (i. e., the spatial scales along the outflow axis) range from 1 to 
18$^{\prime\prime}$, while the $a_{x,k}$ (i. e., the spatial scales across the outflow axis) 
have values of up to 15$^{\prime\prime}$. 

For knots I-N, the $a_{x,k}$ values are either smaller than 6$^{\prime\prime}$ or larger than 12$^{\prime\prime}$ .  
Interestingly, 
the intermediate scales, which are detected in other regions of the jet, do not appear as local maxima in the wavelet spectrum,
 and are therefore not picked up by the analysis described in \S4.1.  

While basically no correlation is seen between the $a_{y,k}$ and $a_{x,k}$ values in knots A-N, 
at larger distances from the 
source, these values appear to be  
correlated.  For knots O-Q, the points more or less fall on a $a_{y,k}/a_{x,k} = 1.3$ 
line, even though there is a wide scatter for $a_{x,k} < 6^{\prime\prime}$.  
For knots R$_1$-R$_2$, the points roughly fall on 
 $a_{y,k}/a_{x,k} = 0.5$ line, with the exception of few points which show large scales along and across the beam of the jet.  
Therefore, we see a transition from structures that are elongated along the outflow axis 
(for knots O-Q) to transversally oriented structures (for knots R$_1$-R$_2$). There are, however, some elongated structures 
at knots O-Q, and some large structures both across and along the beam of the jet at the location of 
knots R$_1$-R$_2$  (see Fig. 5).    
  Finally, for the S-shaped structure 
(at $y>140^{\prime\prime}$, which we have labeled S) both axially elongated and transversal structures appear to be present.

\section{The spatial distributions of the radial velocities and the line widths}

In order to describe the spatial dependence of the kinematical properties of the HH 110 flow, 
we compute two different moments of the line profiles: 
\begin{equation}
V_c = { \int v I_{v} dv \over \int I_v dv }\,, 
\label{vcd}
\end{equation}
\begin{equation}
W^2 = {\int (v - V_c)^2 I_v dv\over \int I_v dv}\,,
\label{w2d}
\end{equation}
for each spatial pixel of the position-velocity cube.  In these equations, $v$ is the radial velocity, and $I_v$ the intensity (at a fixed position $x,y$) of the successive channel maps.  We carry out the integrals over the radial velocity $v_r$ by replacing them with sums over the velocity channel maps.  $V_c$ (see equation \ref{vcd}) is the barycenter of the line profile (which from now we will call the ``line center'' radial velocity), and $W$ (equation \ref{w2d}) is a second order moment that reflects the width of the line profile.

We find that the value of $W$ is quite strongly affected by the H~II region
(which is particularly bright in the surroundings of the faint S region, see
figure 2). Therefore, we have subtracted this emission by fitting a linear
dependence between the fluxes (seen in each channel map) in two rectangular
regions (with a $10''$ width) on each side of the HH~110 jet.

With the values of $V_c$ and $W$ computed for all positions $(x,y)$ on the plane of the sky, we then compute the following spatial averages: 
\begin{equation}
< W^2 > = {\int_{S_{a_x, a_y}} W^2 (x^\prime, y^\prime) I(x^\prime, y^\prime) dx^\prime dy^\prime \over \int_{S_{a_x, a_y}} I (x^\prime, y^\prime) dx^\prime dy^\prime} , 
\label{wavd}
\end{equation}
\begin{equation}
<V_c> \, = \, {\int_{S_{ a_x, a_y}} V_c (x^\prime, y^\prime) I (x^\prime, y^\prime) dx^\prime dy^\prime \over \int_{S_{a_x, a_y}} I (x^\prime, y^\prime) dx^\prime dy^\prime} , 
\label{vavd}
\end{equation}
\begin{equation}
< \Delta v^2 > = { \int_{S_{a_x, a_y}} \left[V_c (x^\prime, y^\prime) - < V_c > ( x, y)\right]^2 I (x^\prime, y^\prime) dx^\prime dy^\prime \over \int_{S_{a_x, a_y}} I (x^\prime, y^\prime) dx^\prime dy^\prime} , 
\label{dvavd}
\end{equation}
where $I(x^\prime, y^\prime)$ is the H$\alpha$ flux obtained from co-adding all of the channel maps.  These integrals are carried out over areas $S_{a_x, a_y}$ which are ellipses with central positions $(x,y)$ and major/minor axes $a_x$ and $a_y$ corresponding to all of the wavelets that have been identified as peaks of the wavelet spectrum (see \S4.1 and figure 3). $<W^2>^{1/2} \, (x,y)$ corresponds to the line width spatially averaged over the ellipse $S_{a_x, a_y}$ with a weight $I(x^\prime, y^\prime)$ (see equation \ref{wavd}).  We then compute the average of the line center velocities $<V_c>$ within the ellipse, as well as the standard deviation $< \Delta v^2 > ^{1/2}$ of these velocities (see equations \ref{vavd} and \ref{dvavd}).  Such deviations of the line center velocity, averaged over sizes chosen from a wavelet spectrum, have been previously used by Gill \& Henriksen (1990) to study turbulence in molecular clouds. 

We then obtain values for $< W^2>^{1/2}$, $<V_c>$ and $< \Delta v^2 >^{1/2}$ at all of the central points of the wavelets shown in Figure 3 (corresponding to peaks in the wavelet spectrum).  In Figure 6, we show the line centers, widths and standard deviations as a function of position $y$ along the jet (all of the points at different positions $x$ across the jet beam and with different $a_x$ and $a_y$ are plotted). 

$<V_c>$ as a function of position $y$ shows quite a wide scatter, and a general trend of more negative velocities 
from knot A ($<V_c>\sim -5$ km s$^{-1}$) to knots P-Q $(< V_c  > \sim - 35$ km s$^{-1}$).  If we measure these 
radial velocities with respect to the surrounding molecular cloud (with a heliocentric radial velocity of +23 km s$^{-1}$, 
see Reipurth \& Olberg 1991) and correct for an orientation angle of 35$^\circ$ with respect to the plane 
of the sky (Riera et al. 2003), we then obtain that the (de-projected) jet velocity grows from $\sim 50$ 
to $\sim$ 100 km s$^{-1}$ from knot A out to knots P-Q.  This is in good agreement with the results obtained 
from long-slit spectra by Riera et al. (2003).

At larger distances $(y > 120^{\prime\prime})$, $< V_c>$ grows, reaching $\approx 0$ at $y = 145^{\prime\prime}$, 
and then remaining with slightly negative values up to the end of the detected emission.  We believe that this 
behaviour of $< V_c>$ for $y > 120^{\prime\prime}$ could be a result of changes in the orientation angle of the 
outflow with respect to the plane of the sky (because in this region, substantial side-to-side excursions of 
the jet beam are observed on the plane of the sky, see Figure 3), and might therefore not reflect a real slowing down of the jet flow. 

The standard deviation $< \Delta v^2 > ^{1/2}$ of the line center velocity (see Figure 6) appears not to have any dependence on position $y$ along the jet beam (showing values mostly within the 7 to 15 km s$^{-1}$ range) out to knot $\hbox{M} \, (y \sim 75^{\prime\prime})$.  Beyond knot N
$(y > 90^{\prime\prime})$, $< \Delta v^2 > ^{1/2}$ shows a more or less monotonic growth, reaching a value of $\sim 25$ km s$^{-1}$ at $y = 175^{\prime\prime}$. 

The average line widths $<W^2>^{1/2}$ do not show a systematic trend with position along the jet, mostly staying within a 
range from 35 to 45 km s$^{-1}$.  This width of course includes the instrumental profile, which has a 
value of $\approx 32$ km s$^{-1}$. 

An interesting feature is that many of the points in the $<V_c>$,
$<\Delta v^2>^{1/2}$ and $<W^2>^{1/2}$ vs. $y$ plots (see figure 3) appear
to fall on more or less monotonic curves. This is somewhat surprising
given the complex structure of the HH~110 outflow. As might be expected, the
points which fall on the monotonic curves are the spatial averages
corresponding to the wavelets with larger sizes ($a_x,\,a_y\sim 10''$, see figures
4 and 5), which naturally give a smoother dependence of the average
quantities as a function of position along the jet.

The fact that the
averages over sizes of $\sim 10''$ give smooth curves as a function of
position indicates that the kinematical properties of the flow have
structures only in sizes $<10''$, which are superimposed on a smooth
dependence on position along the jet axis. To some extent, this result
is a quantification of the fact that while the H$\alpha$ intensity
map of HH~110 has complex, high contrast
structures with many different spatial scales,
the radial velocity and line width maps of this object only show small scale
structures which are superimposed on a generally smooth distribution as
a function of position along the jet (see figure 2).

In Figure 7, we plot the line centers, widths and standard deviations as a function of position $x$ 
across the jet (all of the points with different $a_x$ and $a_y$ are plotted).  We plotted together 
the points with values of $y$ in five different intervals along the jet beam (given in Table~2).

$<V_c>$ shows basically flat structures as a function of $x$ (i. e., across the jet beam) in the regions of knots A to N.   
For the three regions at larger distances from the source (O-Q,  R$_1$-R$_2$ and S), either growing or 
decreasing $<V_c>$ vs. $x$ trends are seen.  The average line width $< W^2 >^{1/2}$ 
shows basically no trend as a function of $x$. 

For knots A-N, the dispersion of the line center $< \Delta v^2 > ^{1/2}$ shows a central region 
(with $| x| < 6^{\prime\prime})$ with values ranging from zero to 12 km s$^{-1}$, and two 
outer peaks (with $|x| = 6 \to 10^{\prime\prime})$ with larger velocity dispersions of up to $\sim 20$ km s$^{-1}$.  
The central, low velocity dispersion region therefore has a diameter of $\sim 12^{\prime\prime}$ 
($8 \times 10^{16}$ cm at 460 pc), and is surrounded by a high velocity dispersion 
envelope with an outer diameter of $\sim 16^{\prime\prime}$ ($1.1 \times 10^{17}$ cm).

Finally, in Figure 8 we show the values of $< \Delta v^2>^{1/2}$ and $< W^2 >^{1/2}$ as a function of the size $a_x$ (across the jet beam) of the ellipses over which the averages have been computed.  Again, we divide the points into the five chosen regions along the outflow axis. 

From Figure 8, it is clear  that both $<\Delta v^2>^{1/2}$ and $< W^2>^{1/2}$ show very clear trends as a function of the size scale $a_x$. The $<W^2>^{1/2}$ values are approximately constant as a function of $a_x$, and show a decreasing scatter for larger $a_x$.  The $<\Delta v^2>^{1/2}$ values also show a decreasing scatter as a function of $a_x$, but this decrease in scatter is superimposed on a shallow trend of increasing $< \Delta v^2 >^{1/2}$ vs. $a_x$. 

Interestingly, in this representation (of $< \Delta v^2 >^{1/2}$ and $< W^2 >^{1/2}$ vs. $a_x$), the line widths and central velocity variations show a quite systematic behaviour.  Because of this, we choose this particular representation in order to try to interpret the kinematical characteristics of the HH 110 jet.  This is done in the following section. 

\section{A turbulent wake (or jet) model}

It is not inmediately clear what the observed line widths and radial
velocity dispersions as a function of spatial scale are actually telling
us.  In order to interpret these results, we use a simple, analytic
model (described in Cant\'o, Raga \& Riera 2003), which allows at least a partial understanding of the implications
of the observations.  This model is as follows. 

It is known from laboratory experiments that the mean flow velocity
of a turbulent jet or wake has a centrally peaked cross section. Superimposed
on this flow are the turbulent fluctuations, which cannot be modeled
in detail.

HH 110 has a complex structure, probably formed by irregularly shaped
shocks which give rise to the observed emission line knots.  The
observed spatially resolved line profiles are then the result of a
mean motion along the jet flow (the ``mean flow'') combined with the
scattered motions resulting from the non-planar
shocks (the ``turbulent velocities'', in analogy to the standard
description of a laboratory jet).  The distribution of the radial
velocity (projected along the line of sight) over the cross section of
the jet will then be determined by the appropriately projected
distributions of the mean and turbulent velocities.

As a simple parametrization of the mean flow velocity, we assume that
it is directed along the outflow axis, and that it has a dependence
on the cylindrical radius $r$ which is given by a generic Taylor series
expansion of the form $v_j = a + br + cr^2 + \dots$.
One can argue that the first order term should be nill, since it leads to
an unphysical on-axis ``cusp'' in the distribution.
Therefore, the lowest order function that we can consider is quadratic.
We will then assume that the radial velocity cross section of the jet
is given by 
\begin{equation}
v_j (r) = v_0 \left( 1 - {r^2 \over h^2} \right)\,,
\end{equation}
where $v_0$ is the central velocity and $h$ the outer radius of
the jet beam. It is assumed that both $v_0$ and $h$ vary along
the jet beam only in scales much larger than the jet width.

If the jet axis lies at an angle $\phi$ with respect to the plane
of the sky (moving away from the observer), the profile of an emission
line for a point within the jet is given by~:
\begin{equation}
j_v={j_0\over {\sqrt{\pi}\Delta v_T}}e^{-[(v-v_r)/\Delta v_T]^2}\,,
\label{jv}
\end{equation}
where
\begin{equation}
v_r=v_0\sin\phi\left[1-{{x^2+z^2\cos^2\phi}\over h^2}\right]\,.
\label{vr}
\end{equation}
In this equation, $x$ is measured across the section
of the jet, and $z$ along the line of sight (with $x,z=0,0$ on the
outflow axis). The line width $\Delta v_T$ includes both the thermal
width and the turbulent velocity (the instrumental width could be
added as well to this term in order to reproduce observed results),
and is assumed to be independent of position within the jet. We also
assume that the emitted energy per unit time, volume and solid
angle in the line $j_0$ is independent of position (see
Cant\'o, Raga \& Riera 2003).

The line intensity at a distance $x$ from the jet axis will then be
\begin{equation}
I_v(x)=2\int_0^{z_m}j_v\,dz\,,
\label{iv}
\end{equation}
where
\begin{equation}
z_m={{\left(h^2-x^2\right)^{1/2}}\over \cos \phi}\,.
\label{zm}
\end{equation}

One can then compute the barycenter velocity
\begin{equation}
V_c (x) = {2\over 3} v_0 \sin \phi\,\left( 1 - {x^2\over h^2} \right)\,, 
\label{vc}
\end{equation}
and the line width
\begin{equation}
W^2(x) = {\Delta v_T^2 \over 2}+{4\over 45} v_0^2 \sin^2\phi \left( 1 - {x^2 \over h^{ \,\, 2}_x} \right)^2\,,
\label{w2a}
\end{equation}
by carrying out the integrals of equations (\ref{vcd}) and (\ref{w2d}).

Now, from our Fabry-P\'erot observations we have computed averages of
the line width over regions of different characteristic sizes $a$
across the width of the jet. In our model, this can be computed as 
\begin{equation}
<W^2> (a,x) = {1\over 2\,a} \int^{x + a}_{x - a} w^2(x^\prime) dx^\prime =
{\Delta {v_T}^2\over 2}+
{4\over 45} {v_0}^2\sin^2\phi\,
\left[ {1\over 5}{a^4\over h^4}-{2\over 3}{a^2\over h^2}
\left(1-3{x^2\over h^2}\right)+\left(1-{x^2\over h^2}\right)^2\right]\,.
\label{w2}
\end{equation}
We can also compute the spatially averaged line center velocity
\begin{equation}
<V_c>(a,x)={1\over 2\,a} \int^{x + a}_{x - a} V_c(x^\prime) dx^\prime =
{2\over 3} v_0\sin\phi\,\left[1-{x^2\over h^2}-{a^2\over 3h^2}\right]\,,
\label{vcav}
\end{equation}
as well as the square of the standard deviation of the line center velocities
\begin{equation}
<\Delta v^2>(a,x)={1\over 2\,a} \int^{x + a}_{x - a}
\left[V_c(x')-<V_c>(a,x)\right]^2\,dx^\prime=
{16\over 27}{v_0}^2\sin^2\phi\,\left({a\over h}\right)^2
\left[{x^2\over h^2}+{a^2\over 15h^2}\right]\,.
\label{dv2}
\end{equation}
Equations (\ref{w2}) and (\ref{dv2}) define an upper envelope and
a lower envelope for the possible values that $<W^2>$ and $<\Delta v^2>$
can take as a function of the averaging scale $a$. These envelopes
are obtained by setting $x=0$ (averaging interval centered on the
jet axis), and $x=h-a$ (averaging interval touching the outer edge
of the jet beam). Setting $x=0$, we obtain a lower envelope for
$<W^2>^{1/2}$ and an upper envelope for $<\Delta v^2>^{1/2}$~:
\begin{equation}
{<W^2>^{1/2}}_{low}(a)=\left\{{\Delta {v_T}^2\over 2}+
{4\over 45} {v_0}^2\sin^2\phi\,
\left[ {1\over 5}{a^4\over h^4}-{2\over 3}{a^2\over h^2}+
1\right]\right\}^{1/2}\,,
\label{wmin}
\end{equation}
\begin{equation}
{<\Delta v^2>^{1/2}}_{up}(a)={4\over 9\sqrt{5}}v_0\sin\phi
\left({a\over h}\right)^2\,.
\label{dvmax}
\end{equation}
Setting $x=h-a$, we obtain an upper envelope for
$<W^2>^{1/2}$ and a lower envelope for $<\Delta v^2>^{1/2}$~:
\begin{equation}
{<W^2>^{1/2}}_{up}(a)=\left\{{\Delta {v_T}^2\over 2}+
{64\over 45} {v_0}^2\sin^2\phi\,\left({a\over h}\right)^2
\left[ {1\over 5}{a^2\over h^2}-{1\over 2}{a\over h}+
{1\over 3}\right]\right\}^{1/2}\,,
\label{wmax}
\end{equation}
\begin{equation}
{<\Delta v^2>^{1/2}}_{low}(a)={4\over 9\sqrt{5}}v_0\sin\phi
\left({a\over h}\right)\left[\left({a\over h}\right)^2+15
\left(1-{a\over h}\right)^2\right]^{1/2}\,.
\label{dvmin}
\end{equation}
The curves for ${<W^2>^{1/2}}_{low}$ and ${<W^2>^{1/2}}_{up}$ 
(as well as the ones for ${<\Delta v^2>^{1/2}}_{low}$ and
${<\Delta v^2>^{1/2}}_{up}$) meet at the maximum possible value
for the averaging length $a$ (i.~e., at $a=h$).

We have drawn the curves given by equations (\ref{wmin}-\ref{dvmin})
on the corresponding plots of Figure 8. For each section along
the jet we have chosen values of $h$ and $v_0\sin\phi$ such
that for $a=h$ (when the upper and lower envelopes join,
see above) the model coincides with the value of
$<\Delta v^2>^{1/2}$ corresponding to the largest measured
scale length perpendicular to the HH~110 axis, which has
$a_x=a_{x,max}$ and $<\Delta v^2>^{1/2}={<\Delta v^2>_{max}}^{1/2}$.
In other words, we choose $h=a_{x,max}$ and
$v_0\sin\phi=(9/4)\sqrt{5}{<\Delta v^2>_{max}}^{1/2}$.
We then choose a value for $\Delta v_T$ so that for $a=h$
the model prediction coincides with the outer point
of the observed $<W^2>^{1/2}$ vs. $a$ points.
The values of $h$, $v_0\sin\phi$ and $\Delta v_T$ resulting
from these fits to the different regions along the HH~110
jet are given in Table~2.

Interestingly, we again find an acceleration along the HH~110 jet,
with $v_0$ increasing from 113~km~s$^{-1}$ (in the A-H region) to
200~km~s$^{-1}$ (in the S region). We can use the obtained
values of $v_0\sin\phi$ to calculate the maximum on-axis value
$<V_c>_{max}=(2/3)v_0\sin\phi$ (obtained by setting $x=0$ and $a=0$
in equation \ref{vcav}). The resulting values (which are also
given in Table 2) should be compared with the observations
by looking at the on-axis values of $<V_c>_{axis}$ in Figure~7.
From this figure, we obtain $<V_c>_{axis}=-12$, $-15$, $-22$, $-18$ and
$-7$~km~s$^{-1}$ for the five chosen regions. If we consider the absolute values of the
corresponding velocities with respect to the cloud (which has
$V_{hel}=+23$~km~s$^{-1}$), we obtain $|<V_c>_{axis}-V_{hel}|=
35$, $38$, $45$, $41$ and $30$~km~s$^{-1}$. For knots A-Q, these values
lie within 20~\%~ of the $<V_c>_{max}$ values obtained from the
fits to the $<\Delta v^2>^{1/2}$ plots (see figure 8 and table 2).

Also interesting is the fact that in order to fit the oberved
$<W^2>^{1/2}(a_x)$ values, we need to include a line broadening
of $\Delta v_T\approx 45$~km~s$^{-1}$. If we subtract in quadrature
the 32~km~s$^{-1}$ instrumental resolution, we obtain an intrinsic
broadening $\Delta v_{T,int}=35\to 46$~km~s$^{-1}$. These values
are substantially larger than the $\sim 10$~km~s$^{-1}$ expected
thermal width, so that they therefore probably reflect turbulent
motions in the jet flow.

Looking at Figure 8, we see that the observed $<W^2>^{1/2}$
and $<\Delta v^2>^{1/2}$ mostly lie on the model predictions for averaging
intervals centered on the outflow axis (equations \ref{wmin} and
\ref{dvmax}, respectively). This is not surprising since the
wavelets chosen by our knot detection algorithm mostly lie close
to the HH~110 axis (see figure 3).

However, for $a_x<5''$ we see that the observed $<W^2>^{1/2}$
and $<\Delta v^2>^{1/2}$ values lie substantially above the
theoretical curves. In particular, the points corresponding to
regions that lie further
away from the outflow axis (shown with open circles in figure 8)
are the ones that tend to have larger velocity dispersions. These
points can be clearly seen to lie in
the outer, high velocity dispersion envelope which is seen
in the cross sections shown in Figure~7 (also see section 5).
We speculate that this envelope corresponds to a turbulent boundary
layer (which is not present in our simple, analytic jet model)
with higher turbulent velocities than the central region of the jet
beam.

\section{Conclusions}

We have obtained a Fabry-P\'erot position-velocity H$\alpha$ cube
of the HH~110 flow. Both the velocity channel maps and the integrated
H$\alpha$ image show a complex structure of knots distributed
in a broadening, more or less conical structure.

Given the
description-defying complexity of the structure, we have chosen
to carry out an anisotropic wavelet analysis of the H$\alpha$
image, which automatically detects the position and characteristic
sizes (along and across the jet axis) of the knots.
From this analysis, we find that the jet first opens out in
a cone with a half-opening angle of $\sim 10^\circ$ for distances up to 
 $70''$ from knot A, and has a constant width out to knot R.  

We have then computed the central velocity, the velocity width and the
dispersion of the central velocity of the line profiles, averaged
over ellipses corresponding to the sizes of the structures detected
by the wavelet analysis. These quantities show that while there is an
acceleration of the jet out from knot A to knot Q (in the region
within $120''$ from knot A), the line width and central velocity
dispersion remain relatively constant. A more complex behaviour
is seen for knots R-S, which could be a reflection of the fact
that in this region the jet has a curved structure (in the plane
of the sky, and therefore probably also along the line of sight).

Across the beam of the jet, we see that there is a central, low
velocity dispersion region with a diameter of $\sim 12''$,
surrounded by a higher velocity dispersion envelope with an
outer diameter of $\sim 16''$. This result could be evidence
for the existence of a turbulent boundary layer in the outer regions
of the HH~110 jet beam.

Finally, for 5 different regions along the jet beam,
we have plotted the mean velocity width $<W^2>^{1/2}$ and central velocity
dispersion $<\Delta v^2>^{1/2}$
as a function of the size scale $a_x$ (perpendicular to
the jet axis) of the detected structures. We find that these
plots show well organized behaviours.

Interestingly, we find that it is possible to model the
observed $<W^2>^{1/2}(a_x)$ and $<\Delta v^2>^{1/2}(a_x)$ profiles
with a simple, analytic model of a
turbulent jet (or wake). From a comparison
of the models with the data, we deduce that most of the observed
values can be explained with a model with an axially peaked mean
flow velocity, with superimposed turbulent (+thermal) motions with a velocity
dispersion of $\sim 40$~km~s$^{-1}$.

However, for characteristic sizes $<5''$, the observed velocity dispersions
lie up to $\sim 25$~km~s$^{-1}$ above the model prediction. We find that
these points correspond to the outer, high velocity dispersion envelope
of the HH~110 jet (see above). Therefore, in this representation we
again see the presence of what appears to be a turbulent boundary
layer, which is not present in our analytic turbulent jet/wake model.

Finally, we should note that the acceleration observed along the
HH~110 jet does not agreee with what one would expect for a turbulent
jet. A turbulent jet incorporates low momentum material from the
surrounding environment, which leads to a progressive decrease
of the flow velocity as one moves downstream along the jet beam
(see, e.~g., Raga et al. 1993). Therefore, the acceleration observed
along the HH~110 jet would favour an interpretation of this flow
in terms of a turbulent wake, rather than a jet. This is the scenario
proposed by Raga et al. (1993), who modeled HH~110 in terms of a wake
left behind by a jet deflected by the surface of
a dense cloud, and then pinched off as the incident jet starts
to burrow into the cloud.

Of course, another possibility would be to have a turbulent jet
ejected with a monotonically increasing initial velocity. If this
increase in the ejection velocity were drastic enough, it could
overcome the drag due to the entrainment of environmental material,
and lead to an ``acceleration'' down the jet axis similar to the
one observed in HH~110.

To conclude, we would like to point out that this work represents
a first attempt to apply wavelet analysis techniques to observations
of astrophysical jets. A spectrum of a 2D image with anysotropic wavelet basis
functions is defined in a four-dimensional space (with two spatial
and two spectral dimensions), and therefore contains a large amount of
information which is difficult to interpret. The present analysis
is based only on a study of the spatial and spectral maxima of the
wavelet spectrum (i.~e., on finding ``knots'' and their characteristic
sizes), and clearly does not use a lot of the information present
in the wavelet spectrum.

\acknowledgments The work of A. Raga and JC was supported by
the CONACyT grant 36572-E and 34566-E. A. Riera acknowledges the
ICN-UNAM for support during her sabbatical. The work of A. Riera was
supported by the MCyT grant AYA2002-00205 (Spain).
BR acknowledges support for this project from NASA grant NAG5-8108 (LTSA) and 
NSF grant AST-9819820.
We acknowledge Lucila Gonz\'alez for her help with the manuscript.

\clearpage

\begin{figure}
\epsscale{0.7}
\plotone{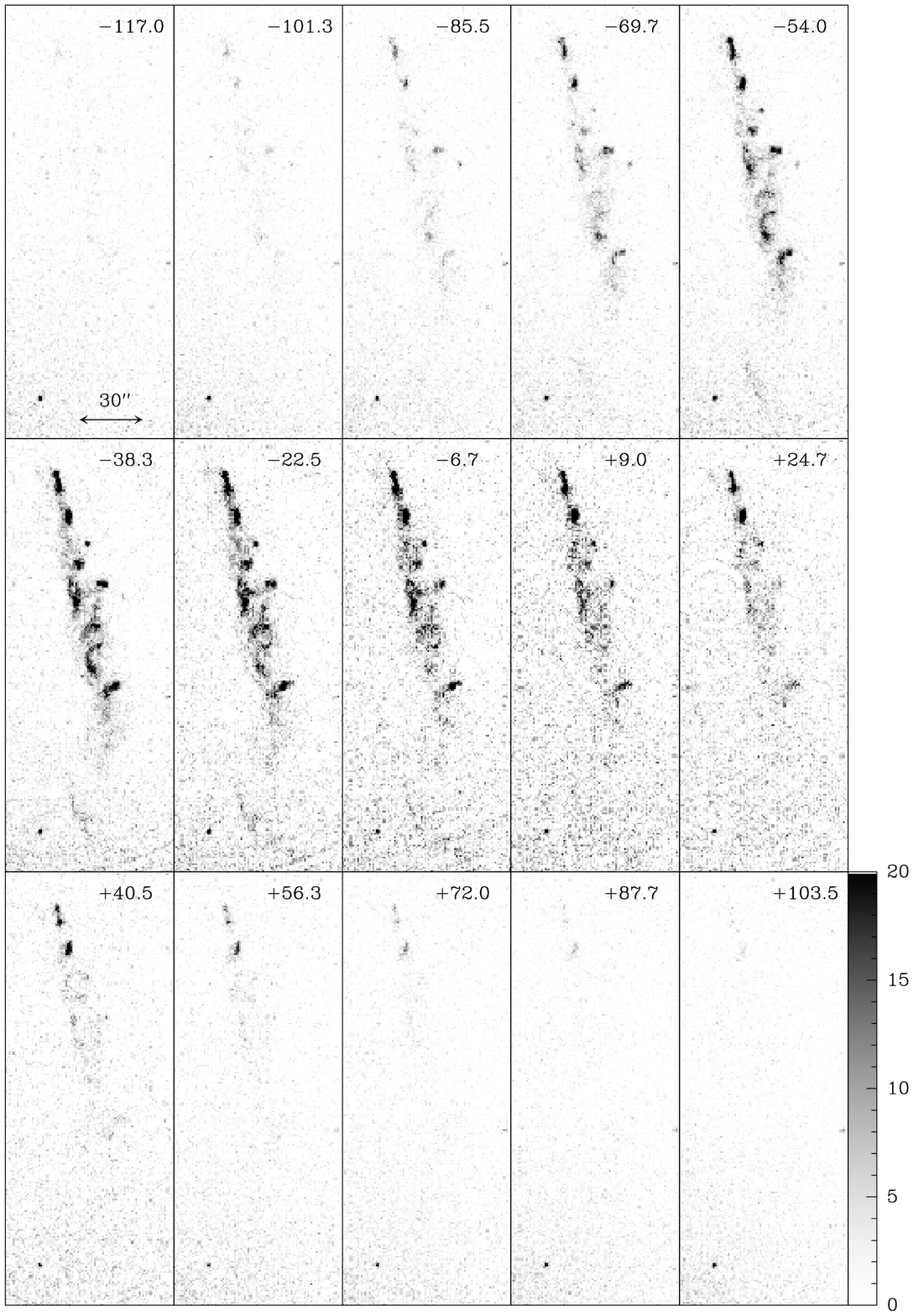}
\caption{H$\alpha$ velocity chanel maps corresponding to various radial velocities
obtained from the Fabry-P\'erot data of HH~110. The heliocentric radial
velocities are indicated. The systemic velocity of the ambient cloud is
$+23$~km~s$^{-1}$. The scale (in arcsec) is given in the top left hand
plot. The maps are shown with a linear greyscale given (in counts) by the
bar.
\label{fig1}}
\end{figure}

\clearpage

\begin{figure}
\plotone{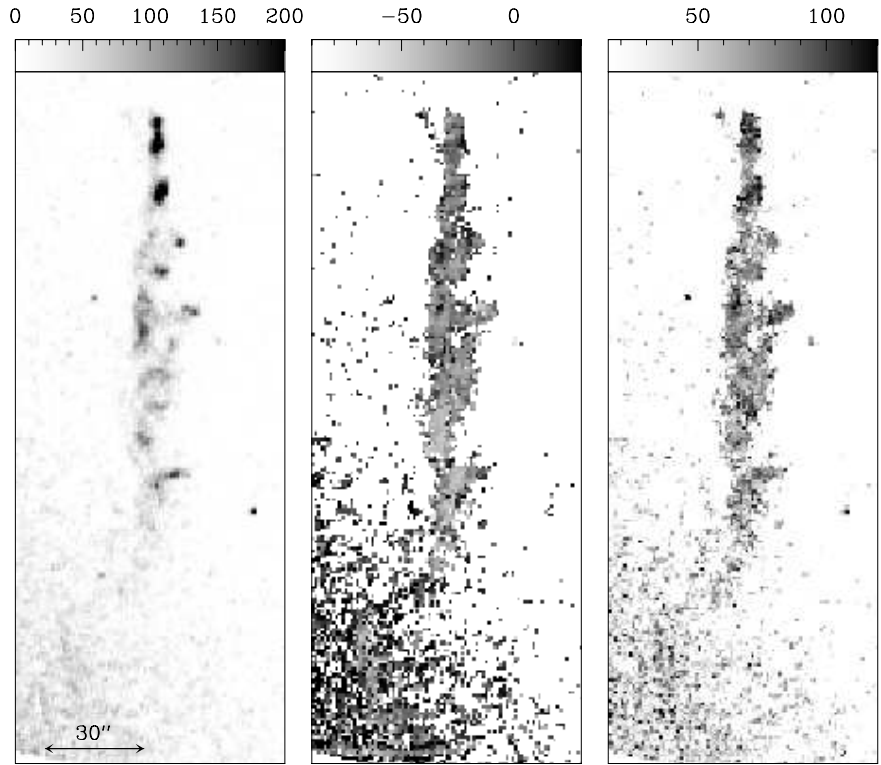}
\caption{The H$\alpha$ image map (left panel), central radial velocity map
(central panel) and 
the line widths (FHWM, right panel) obtained from the position-velocity cube.   
We have rotated the maps by 13$^\circ$ so that the outflow is more or less
parallel to the $y$-axis. The greyscales of the map (linear, in counts) and
of the central radial velocity and line width (linear, in km~s$^{-1}$) are
given by the bars on top of each plot.
\label{fig2}}
\end{figure}

\clearpage

\begin{figure}
\plotone{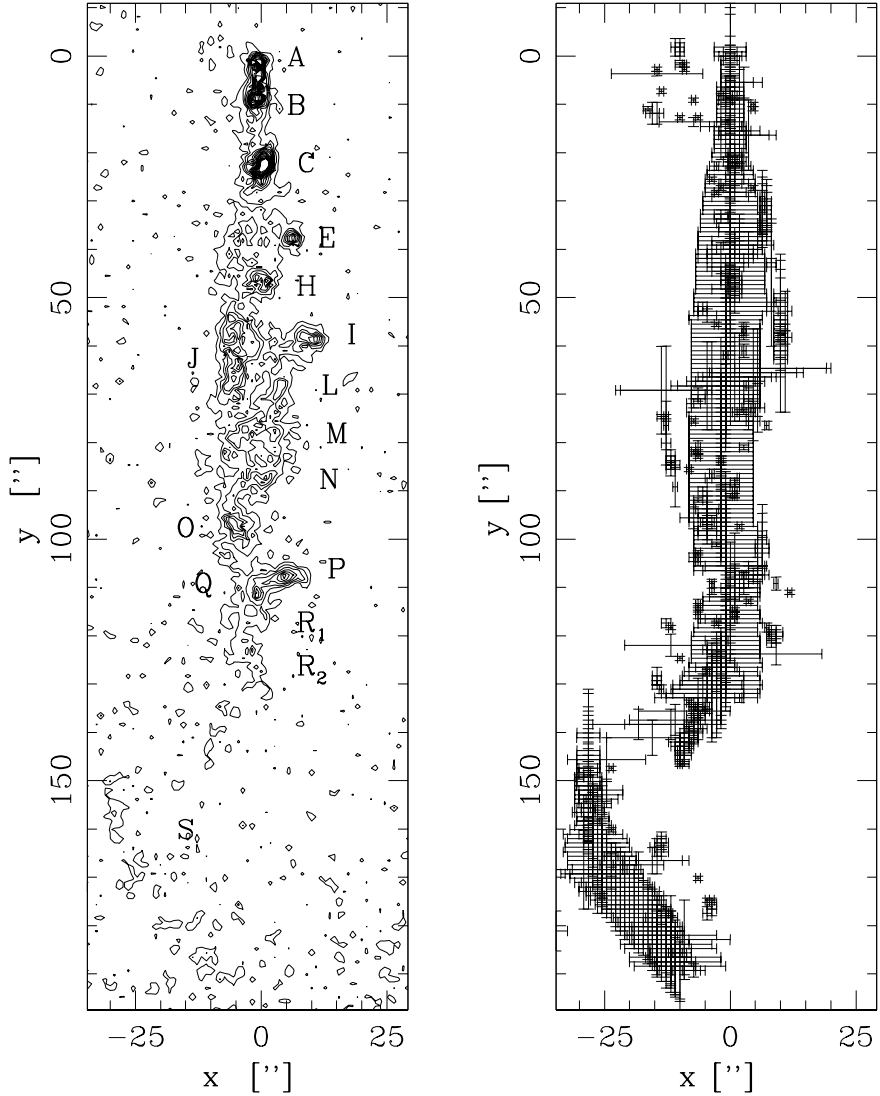}
\caption{Left panel: contour plot of the H$\alpha$ image, where we have
labelled the knots to help with their identification. 
Right panel: the positions and characteristic sizes of the
structures along the HH~110 jet. The crosses show the positions
of the maxima $x_k$ of the wavelet transform obtained for the different
values $y$ along the jet. The characteristic sizes $a_{x,k}$ and
$a_{y,k}$ of these maxima are shown as error bars centered on the
positions of the maxima. The distances
$x$ (across the jet axis) and $y$ (along the jet axis)
are measured from the peak of knot A. We have rotated the maps by
13$^\circ$ so that the outflow is more or less parallel to the $y$-axis.
\label{fig3}}
\end{figure}

\clearpage

\begin{figure}
\epsscale{0.9}
\plotone{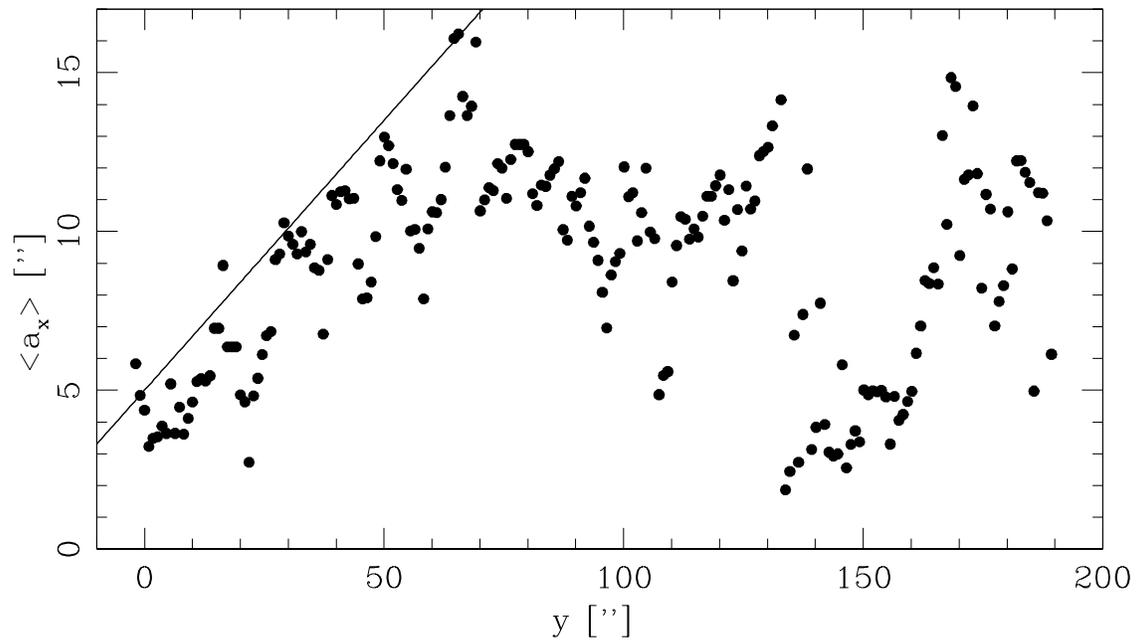}
\caption{The $<a_x>$ are plotted
as a function of position $y$ along the HH~110 jet. A straight line that
forms an upper envelope to the $<a_x>$ values (in the $y<70''$ region)
has been drawn (see the text).
\label{fig4}}
\end{figure}

\clearpage
\begin{figure}
\epsscale{0.33}
\plotone{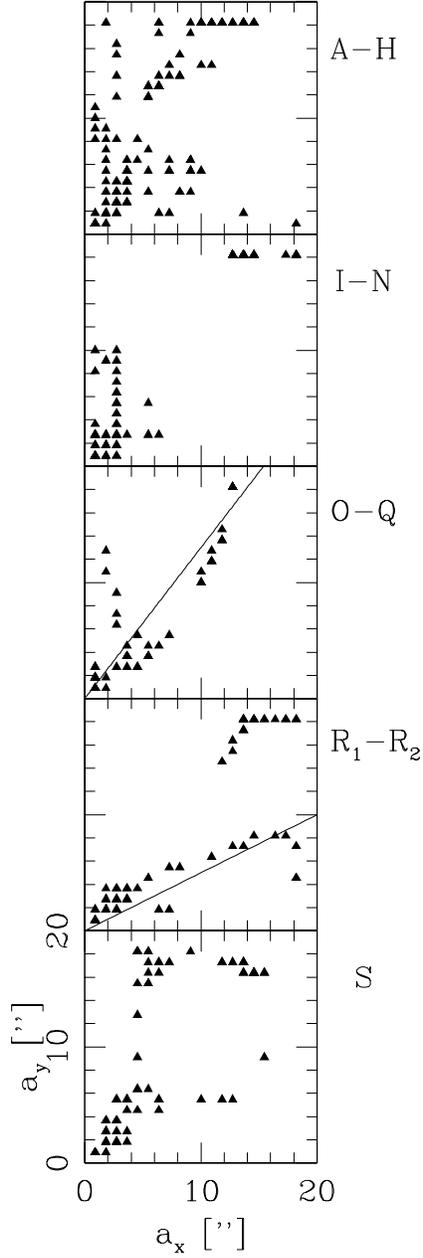}
\caption{The $a_{y,k}$ are plotted as a function of $a_{x,k}$ for different
regions along the HH~110 flow (see Table 2). For regions O-Q and R$_1$-R$_2$ the
results from linear fits to the points are drawn as solid lines. 
\label{fig5}}
\end{figure}

\begin{figure}
\epsscale{1.0}
\plotone{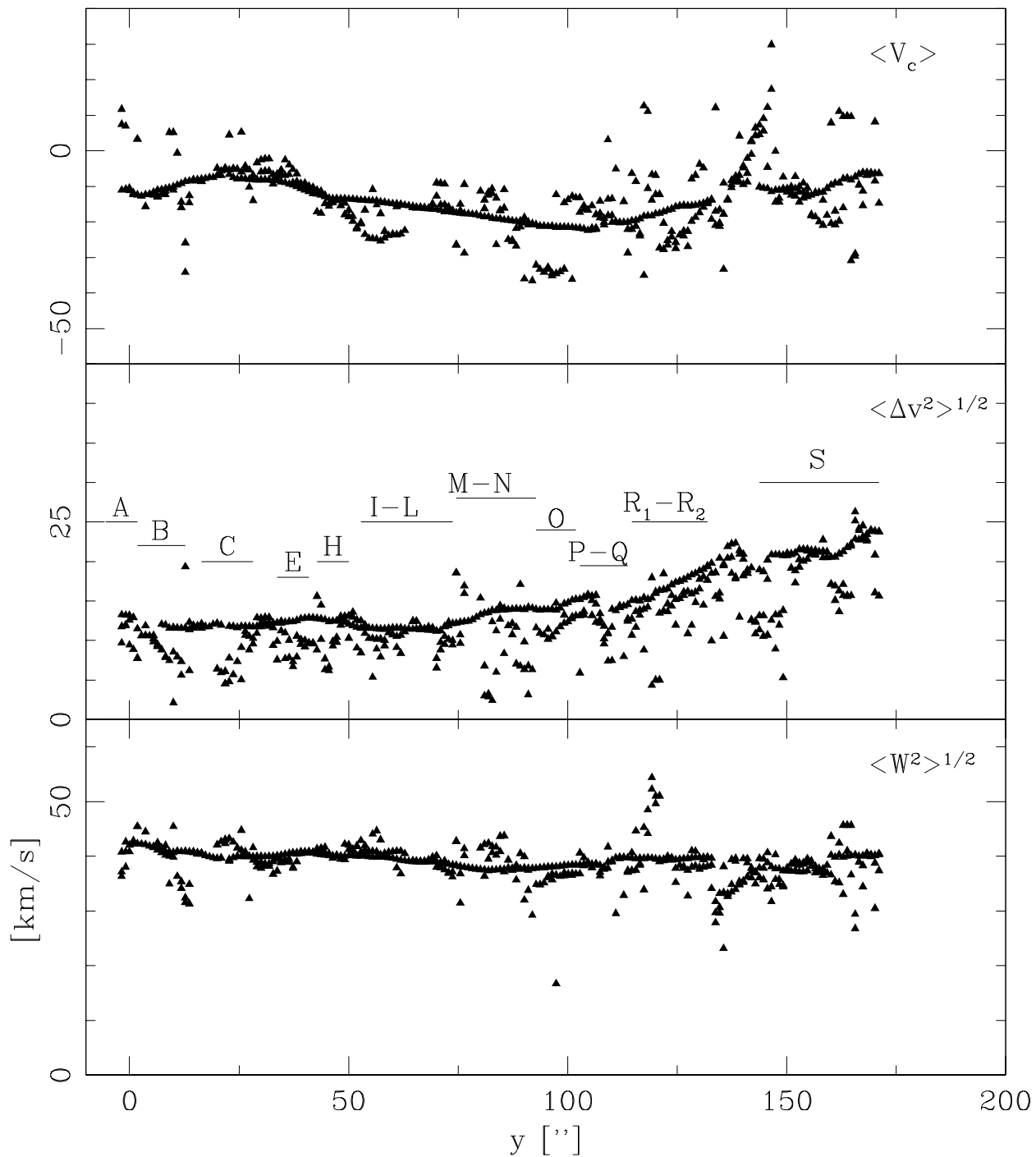}
\caption{The average line centers (top panel), line widths (central panel) and 
standard deviations (bottom panel) as a function of position $y$ along the
HH~110 jet. All points with different positions $x$ across the jet and with
different $a_x$ and $a_y$ are plotted.
\label{fig6}}
\end{figure}

\clearpage

\begin{figure}
\plotone{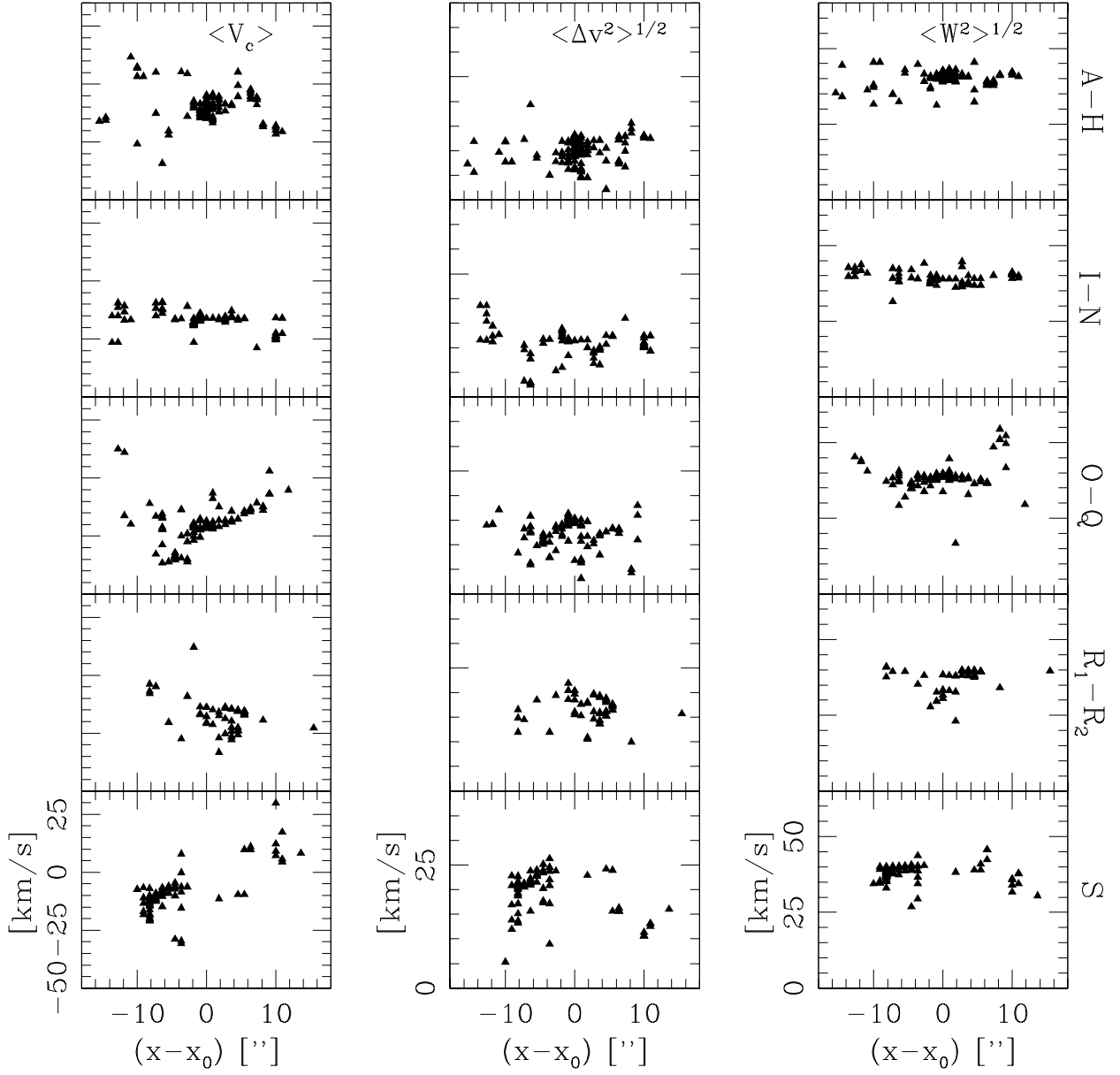}
\caption{The line centers (left), line widths (center) and 
standard deviations (right) are plotted as a function of position
$x$ across the jet beam for the 5 different chosen regions along the HH~110 flow.
$x_0$ is the central position of the jet cross section (which varies along
the jet, as the beam has side-to-side excursions, see figure 3).
\label{fig7}}
\end{figure}

\clearpage

\begin{figure}
\plotone{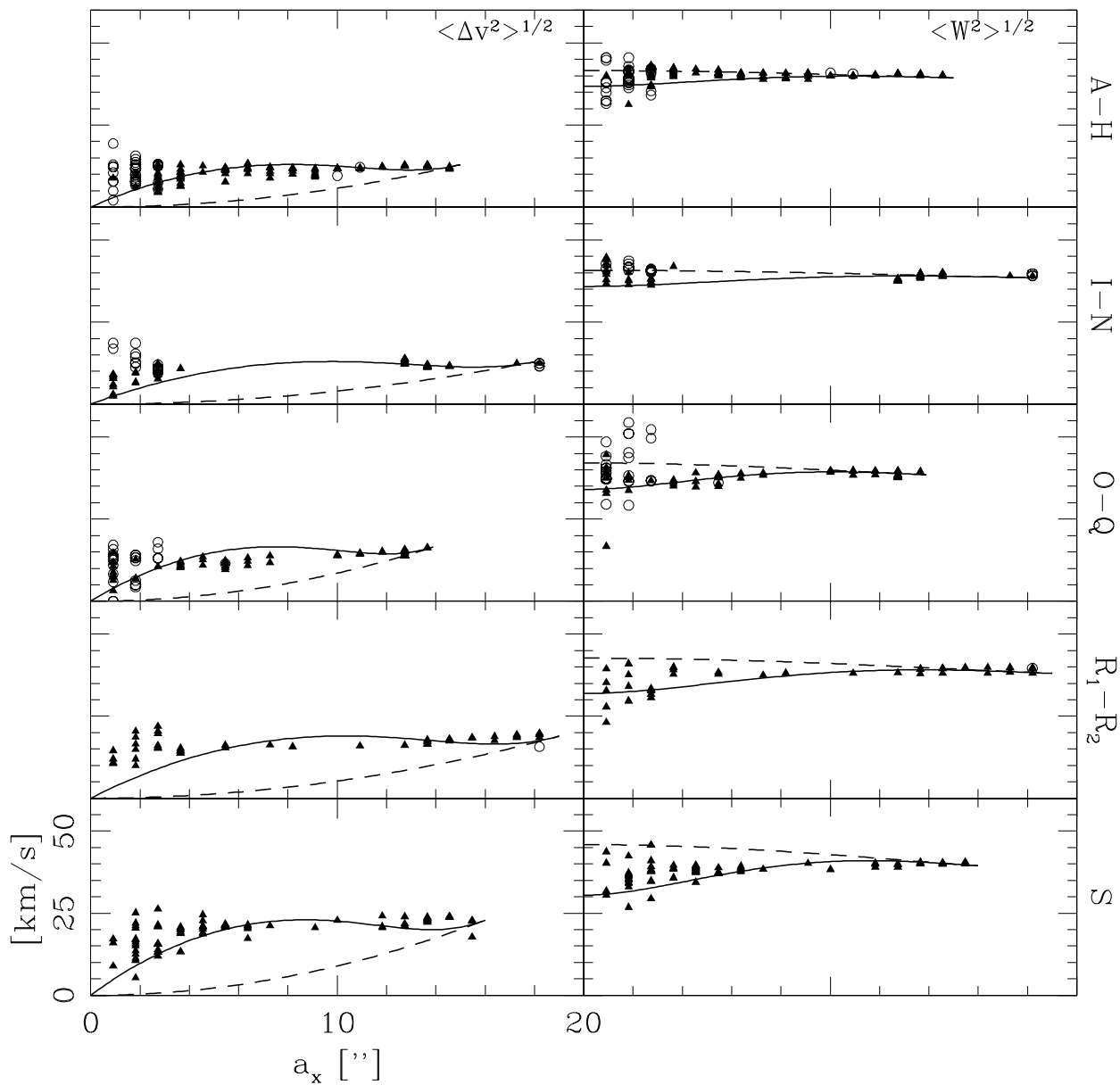}
\caption{The average standard deviations of the line center (left panels) and the
line widths (right panels) obtained from the data cube are plotted as a function of
the size scale $<a_x>$ across the jet flow for the 5 chosen
regions along the jet. The triangles correspond to points which lie in
the central region of the flow (with $|x| \leq 3''$ for the A-H region, 
$|x| \leq 7''$ for the I-N region, $|x| \leq 5\farcs5$ for the O-Q region, and
$|x| \leq 9''$ for the R$_1$-R$_2$  region), and the open dots correspond to the outer
region of the jet beam (see section 6).
The curves correspond to the predictions of the
analytical model, which have been fitted 
independently for each region along the HH~110 jet (see section 6).
In the left panels, the curves correspond to $<\Delta v^2>^{1/2}_{up}$
(solid line) and  $<\Delta v^2>^{1/2}_{low}$ (dashed line). 
In the right panels, the curves correspond 
to $<W^2>^{1/2}_{up}$ (solid line) and $<W^2>^{1/2}_{low}$ (dashed line). 
\label{fig8}}
\end{figure}

\clearpage

\begin{deluxetable}{lll}
\tabletypesize{\scriptsize}
\tablecaption{Journal of Fabry-P\'erot observations}
\tablewidth{0pt}
\tablehead{}
\startdata
Observations & Telescope & ESO 3.6m \\
& Equipment & CIGALE @ Cassegrain \\
& Date & 1997, January, 4-6 \\
& Seeing & $0\farcs4$ $\rightarrow$  $0\farcs8$ \\
Calibration & Neon Comparison light & $\lambda$ 6598.95 \AA \\
Fabry-P\'erot & Interference Order & 793 @ 6562.78 \AA \\
& Free Spectral Range at H$\alpha$ & 380 km s$^{-1}$ \\
 & {\it Finesse}\tablenotemark{a} at H$\alpha$ & 11 \\
& Spectral resolution at H$\alpha$ & 9400\tablenotemark{b} \\
Sampling & Number of Scanning Steps & 24 \\
& Sampling Step & 0.35 \AA (16 km s$^{-1}$) \\
& Total Field & $170'' \times 170'' \, (256 \times 256$ px$^2$) \\
& Pixel Size & 0.91'' \\
Detector & & IPCS (Thomsom tube)\\
\enddata
\tablenotetext{a}{Mean Finesse through the field of view}
\tablenotetext{b}{For a signal to noise ratio of 5 at the sample step}

\end{deluxetable}

\begin{deluxetable}{cccccccc}
\tabletypesize{\scriptsize}
\tablecaption{Model fits to the HH~110 jet}
\tablewidth{0pt}
\tablehead{
\colhead{Region} & \colhead{ $y_{min},\,y_{max}$}   & \colhead{$h$}   &
\colhead{ $v_0 \sin\phi$} &
\colhead{$v_0$}  & \colhead{$<V_c>_{max}$} & \colhead{$\Delta v_T$} &
\colhead{$\Delta v_{T,int}$ }\\
       & $['']$ & $['']$ & \multispan5{\hfil [km~s$^{-1}$]\hfil} \\
}
\startdata
\hline
A-H & \phantom{0}$-11$, 53\phantom{0} & 15 & 65 & 113 & 43 & 52 & 46 \\
I-N & \phantom{$-$0}53, 85\phantom{0} & 18 & 65 & 113 & 43 & 51 & 45 \\
O-Q & \phantom{$-$0}85, 121 & 14 & 83 & 145 & 55 & 48 & 41 \\
R$_1$-R$_2$ & \phantom{$-$}121, 136 & 19 & 95 & 166 & 63 & 45 & 37 \\
S   & \phantom{$-$}144, 171 & 16 & 115 & 200 & 77 & 43 & 35 \\
 \enddata
\tablecomments{$y_{min},\,y_{max}$ give the position of the chosen regions
along the HH~110 jet, where $y$ = 0 corresponds to the peak of knot A.
$h$ is the outer radius of the jet beam and
$v_0 \sin\phi$ is the central radial velocity of the model fit to the
data (see section 6).  The deprojected velocity $v_0$ 
was computed assuming an inclination angle $\phi = 35^\circ$
with respect to the plane of the sky. $<V_c>_{max}$ is the maximum on axis value
for the barycenter of the line profile computed from the model fit.
$\Delta v_T$ is the turbulent+thermal+instrumental line width necessary
to fit the observed $<W^2>^{1/2}$ vs. $a$ (see section 6). 
$\Delta v_{T,int}$ is the intrinsic broadening, computed substracting in 
quadrature the instrumental width from $\Delta v_T$, where $\Delta v_{ins}= 
(FHWM_{ins})/(2 (ln 2)^{1/2})$. 
For an instrumental $FWHM$ = 34 km s$^{-1}$, a $\Delta v_{ins}$ 
= 25 km s$^{-1}$ is obtained. 
}
\end{deluxetable}


\clearpage

\begin{thebibliography}{}

\bibitem[1996]{amr96}Amram P., Balkowski C., Boulesteix J., Cayatte V., Marcelin M. \& Sullivan W., 1996, A\&A, 310, 737

\bibitem[1995]{amr95}Amram P., Boulesteix, J., Marcelin, M., Balkowski, C., Cayatte, V. \& Sullivan, W.T., III, 1995, A\&As, 113, 35

\bibitem[1992]{amr92}Amram, P., Le, Coarer, E., Marcelin, M., Balkowski, C., Sullivan, W. T. III \& Cayatte, V. 1992, A\&As, 94, 175

\bibitem[1994]{amr94}Amram, P., Marcelin, M., Balkowski, C., Cayatte, V., Sullivan, W. T. III \& Le Coarer, E. 1994, A\&As, 103, 5

\bibitem[1998]{amr98}Amram, P., Mendes de Oliveira, C., Boulesteix, J. \& Balkowski, C., 1998, A\&A, 330, 881

\bibitem[1993]{amr93}Amram, P., Sullivan, W., Balkowski, C., Marcelin, M. \& Cayatte, V., 1993, ApJ, 403, L59

\bibitem[Bendjoya(1991)]{bend91}Bendjoya, Ph., Sl\'ezak, E. \& Froeschl\'e, Cl., 1991, A\&A, 251, 312

\bibitem[2003]{Canto03}Cant\'o, J., Raga, A.C. \& Riera, A. 2003, RMxAA (submitted) 

\bibitem[Choi(2001)]{choi01}Choi, M., 2001, ApJ, 550, 817

\bibitem[Davis(1994)]{dav94}Davis, C. J., Mundt, R. \& Eisl\"offel, J., 1994, ApJ, 437, L55

\bibitem[Gill(1990)]{gill90} Gill, A. G. \& Henriksen, R. N., 1990, ApJ, 365, L27

\bibitem[Gouveia(1999)]{gou99}de Gouveia Dal Pino, E. M., 1999, ApJ, 526, 862

\bibitem[Holschneider(1995)]{Hol95}Holschneider, M. 1995, {\it Wavelets : an analysis tool}, Oxford University

\bibitem[Hurka(1999)]{hur99}Hurka, J. D., Schmid-Burgk, J. \& Hardee, P. E., 1999, A\&A, 343, 558


\bibitem[1987]{lav87} Laval, A., Boulesteix, J., Georgelin, Y. P., Georgelin, Y. M. \& Marcelin, M. 1987, A\&A, 175 199


\bibitem[Noriega(1996)]{nor96}Noriega-Crespo, A., Garnavich, P. M., Raga, A. C., Cant\'o, J. \& B\"ohm, K. H., 1996, ApJ, 462, 804

\bibitem[Raga(1995)]{rag95}Raga, A. C., Cant\'o, J., 1995, RMxAA, 31, 51

\bibitem[Raga(1993)]{rag93} Raga, A. C., Cant\'o, J., Calvet, N.,
Rodr\'\i guez, L. F. \& Torrelles, J. M., 1993, A\&A, 276, 539

\bibitem[Raga(2002)]{rag02}Raga, A. C., de Gouveia Dal Pino, E. M., Noriega-Crespo, A., Mininni, P. D. \&
 Vel\'azquez, P. F., 2002, A\&A, 392, 267

\bibitem[Raga(1991)]{rag91}Raga, A. C., Mundt, R. \& Ray, T. P., 1991, A\&A, 246, 535

\bibitem[Reipurth(1991)]{rei91} Reipurth, B. \& Olberg, M., 1991,
A\&A, 246, 535

\bibitem[Reipurth(1996)]{rei96} Reipurth, B., Raga, A. C.\& Heathcote,
S., 1996, A\&A, 311, 989

\bibitem[Riera(2003)]{rie03} Riera, A., L\'opez, R., Raga, A. C.,
Estalella, R. \& Anglada, G., 2003, A\&A, 400, 213

\bibitem[Rodriguez(1998)]{rod98} Rodr\'\i guez, L. F., Reipurth, B.,
Raga, A. C. \& Cant\'o, J., 1998, RMxAA, 34, 69

\end{thebibliography}
\end{document}